\documentclass[11pt,twoside]{article}
\usepackage{asp2006}
\usepackage{psfig}
\usepackage{epsf}
\usepackage{graphics}
\usepackage{lscape}
\newcommand{\gtsim}{\mbox{{\raisebox{-0.4ex}{$\stackrel{>}{{\scriptstyle\sim}}
$}}}}

\def\edcomment#1{\iffalse\marginpar{\raggedright\sl#1\/}\else\relax\fi}
\reversemarginpar

\begin{document}
\title{Science with the next generation of radio surveys from LOFAR to the SKA
}
\author{Matt J. Jarvis}
\affil{Astrophysics, University of Oxford, Denys Wilkinson Building, Keble Road, Oxford, OX1 3RH, UK}

\begin{abstract}
Over the next few years the new radio telescopes, such as the Low
Frequency Array (LOFAR) will greatly enhance our knowledge of the
active history of the Universe. Large-area surveys with these new
telescopes will no longer be dominated by the powerful active galactic
nuclei, but by radio-quiet quasars and star-forming galaxies over all
cosmic epochs.  Further in the future ($\sim 2014$) the Square Kilometre Array (SKA)
will take studies in the radio regime to a whole new level, with the
ability to detect neutral hydrogen via the 21~cm transition over the
majority of cosmic time. This will enable both the detailed study of
individual galaxies and the use of these galaxies as probes Dark
Energy. In these proceedings I give an overview of the science goals behind these new radio telescopes, with particular emphasis on galaxy evolution and cosmology. Finally I briefly discuss the SKA science simulation effort.
\end{abstract}

\vspace{-0.5cm}
\section{Introduction}

During the last half century our knowledge of the Universe has been
revolutionised by the opening of observable windows outside the narrow
visible region of the spectrum. Observations from the radio to
$\gamma$-rays have provided new and completely unexpected information
about the nature and history of the Universe and have resulted in the
discovery of a cosmic zoo of strange and exotic objects.  At this
meeting we have seen the exciting results that are now coming out of
the deep optical and infrared surveys being undertaken from the ground
and space. In this contribution I highlight some of the major projects
in the radio regime that will have the ability to significantly
advance our knowledge of galaxy formation and evolution, in
addition to many other science areas.

\section{The Low-Frequency Array}

One of the
few spectral windows that still remains to be explored is at the low
radio frequencies, the lowest energy extreme of the accessible spectrum.

The Low Frequency Array (LOFAR) is a revolutionary aperture synthesis
array currently under construction in the Netherlands. It will operate
in the $\sim$20--240 MHz (1.25--15m wavelength) range and will achieve at least
4" resolution at 200 MHz with higher angular resolution likely to
materialise as a result of LOFAR being extended into Germany and the
UK. LOFAR will instantaneously survey large areas of the sky using
independently steerable beams. It will be 3 orders of magnitude more
sensitive than previous low-frequency radio telescopes.

One of the most important areas of LOFAR will be to carry out
large-sky surveys. Such surveys are well suited to the characteristics
of LOFAR and have been designated as one of the key projects that have
driven LOFAR since its inception. Such deep LOFAR surveys of the
accessible sky at several frequencies will provide unique catalogues
of radio sources for investigating several fundamental areas of
astrophysics which, up until now, have been within the remit of
telescopes operating at shorter wavelengths. These include tracing the
star-formation history of the Universe, the formation of massive black
holes, galaxies and clusters of galaxies. Moreover, because the LOFAR
surveys will probe unexplored parameter space, it is likely that they
will discover new phenomena.

\subsection{Tracing the history of the active Universe with LOFAR}

As mentioned above LOFAR will have the capability of tracing
star-forming galaxies out to the highest redshifts. This has already
been done with radio telescopes to a certain extent with deep,
small-area ($\sim 1$~square degree) observations with the VLA \citep[e.g.][]{Ivison2002}, however LOFAR
offers the possibility to extend such studies to $\gtsim
100$~square degree areas. This is possible due to the new technology that
LOFAR is built on (see http://www.lofar.org), which enables
multi-beaming over large areas. Thus the survey speed is huge in
comparison to traditional dish-based interferometers operating at
higher frequencies such as the VLA.

The improvement in survey speed is also helped considerably by
operating at low-frequency. This is because the typical power-law
slope of synchrotron radiation emitted from star-forming galaxies and
AGN increases towards low frequencies. Thus, observing at
low-frequencies is important to combat the $k-$correction for the
higher-redshift sources.

\begin{figure}
\plottwo{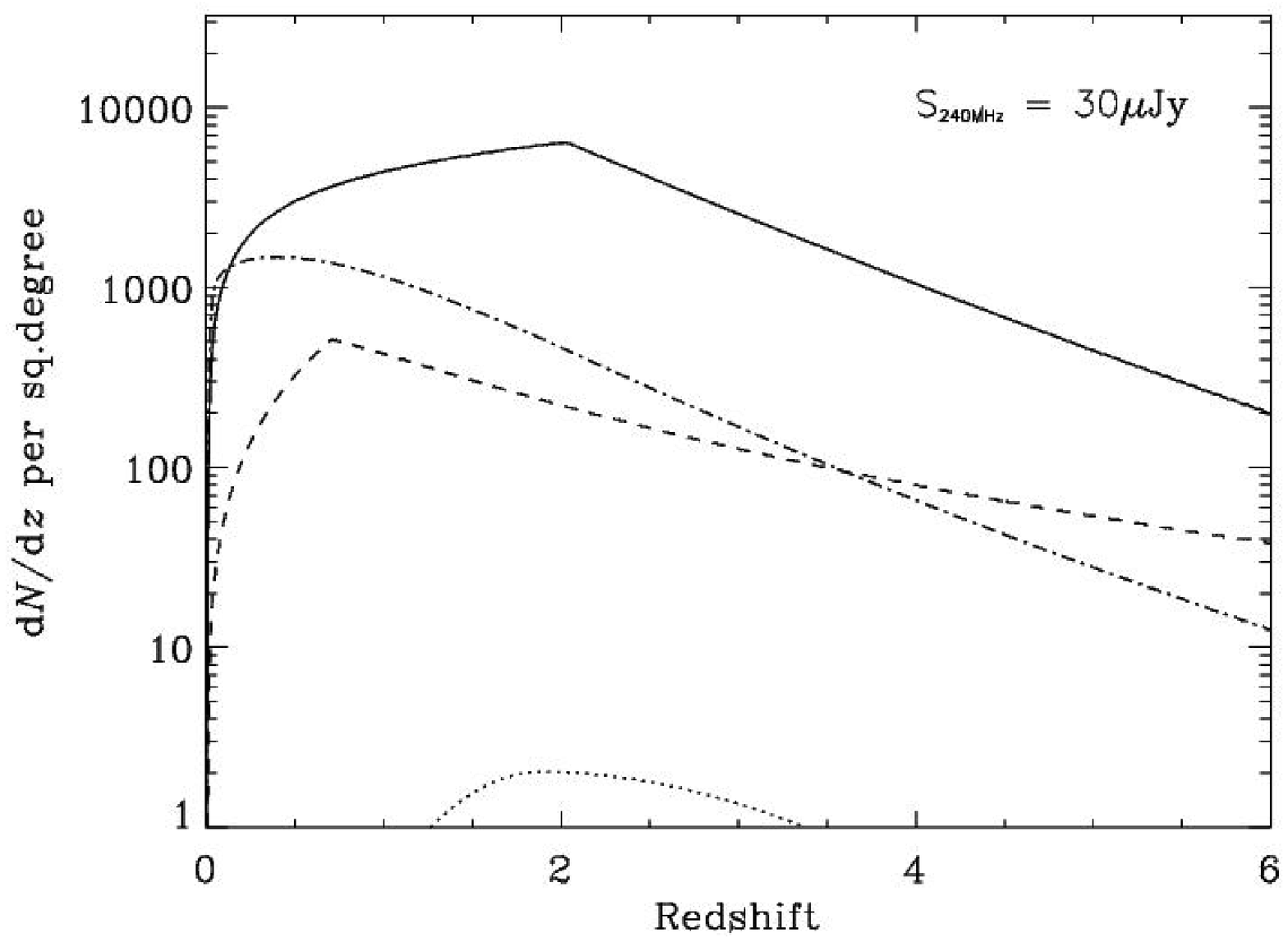}{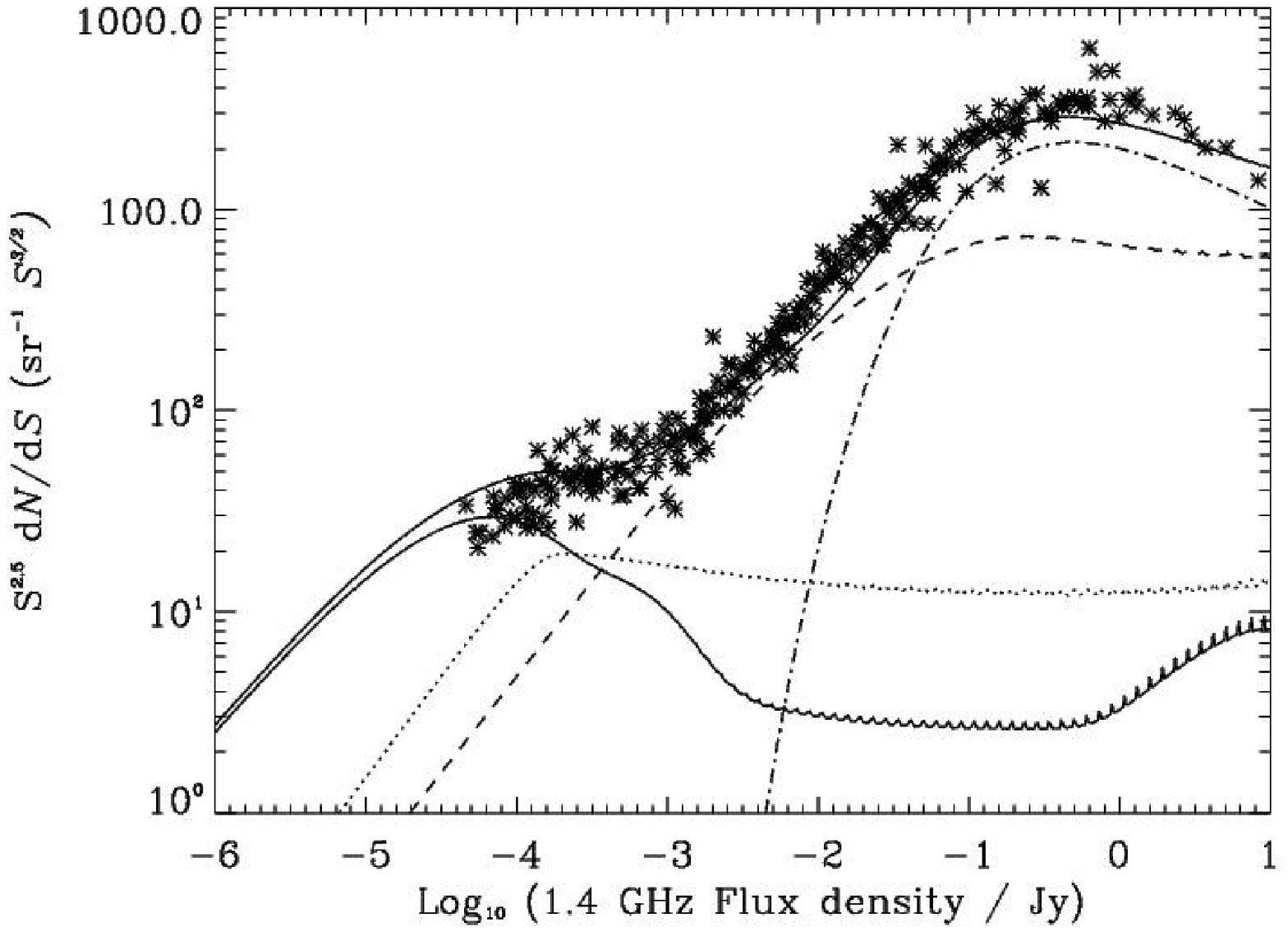}
\caption{{\it (left)} The predicted number of sources per unit redshift
per square degree for a typical deep LOFAR survey with a flux-density
limit of $30~\mu$Jy. The various lines represent star-forming galaxies
(solid), radio-quiet quasars (dot-dashed), FRI-radio galaxies (dashed)
and FRII-radio galaxies (dotted). {\it (right)} The predicted source
counts for these models with the various lines showing the FRIIs
(dot-dashed), FRIs (dashed), radio-quiet quasars (dotted) and
star-forming galaxies (solid). The total source counts are also
represented by the upper solid line. The stars are the 1.4~GHz source
counts from \cite{Seymour04}.}\label{fig:jarvis_dndz}
\end{figure}

To illustrate the massive leap in sensitivity of the LOFAR, we show in Fig.~1
the predicted number of objects per unit redshift per square
degree for a proposed deep-LOFAR survey of 250~square degrees. Details of how
these curves were generated for the AGN populations can be found in
\cite{JR04}, the evolution of starburst galaxies were
modelled using the luminosity function of \cite{Yun01}
with pure luminosity evolution constrained by the far-infrared sources
counts of \cite{Blain99} and the radio
source counts of \cite{Seymour04}, (Jarvis et al. in prep).  One can see that the powerful
active galactic nuclei (AGN) traditionally found in radio surveys are
in the minority in deep LOFAR surveys. Moreover it is also clear that
LOFAR enters the regime where radio-quiet quiet quasars and
low-luminosity Fanaroff-Riley \cite{FR74} class I (FRI) AGN are the dominant AGN
populations over all redshifts, rather than the powerful FRII-type
radio galaxies.  Thus, LOFAR offers the unique opportunity to trace
the majority of activity in the Universe.

However, much of the science that can be carried out with the LOFAR
deep surveys will also rely on data from all of the other
wavebands. Particularly, synergies with the new generation of
sensitive wide-field optical and near-infrared telescopes will be
crucial in identifying the sources responsible for the radio emission,
such as the obscured AGN, naked AGN and starburst galaxies. These will
not only lead to the identification of the radio sources, the
positions of which can subsequently used for spectroscopic follow-up,
but the optical and near-infrared data sets could be used to estimate
redshifts for the various sub-populations. Such a data set over $\sim
250$~square degrees with a high areal density would thus provide the ideal
input catalogue for future experiments to measure the evolution in the
Dark Energy component of the Universe through Baryon Acoustic
Oscillations (BAO) with instruments such as FMOS and WFMOS
\citep[e.g.][]{Nichol06}. Objects selected at radio wavelengths will also be much
easier to obtain redshifts for due to the high probability of them having
bright emission lines from either a starburst or AGN. Therefore, the time to
survey the galaxy populations in spectroscopy is less than the
time that would be needed to obtain redshifts from the continuum
features and absorption lines that are the characteristics of the
massive elliptical galaxies, which are currently being used in various
current experiments \citep[e.g.][]{Cannon06}.

\section{The Square Kilometre Array}

LOFAR will have a collecting area of roughly 5 per cent of a
square-kilometre array (SKA), thus it is a very important path finder
experiment towards the much larger SKA telescope.

\begin{figure}
\plottwo{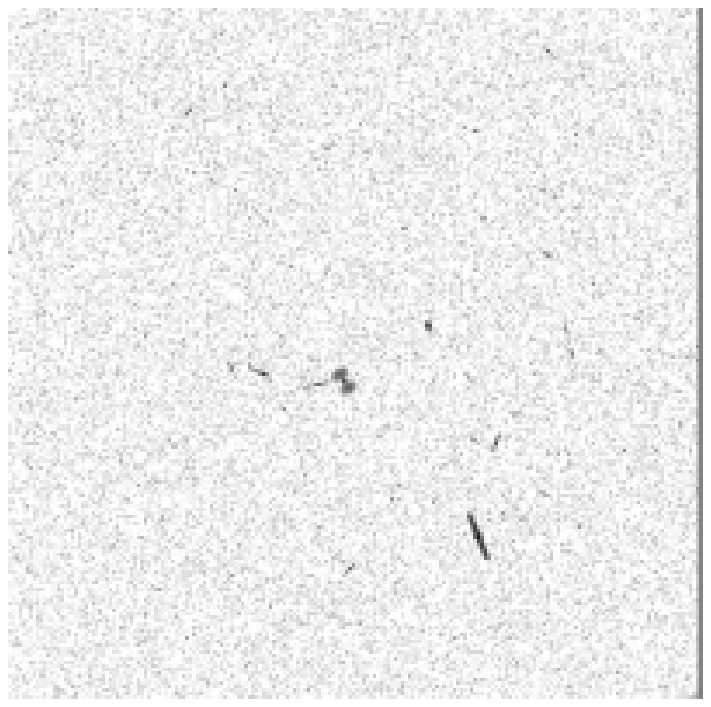}{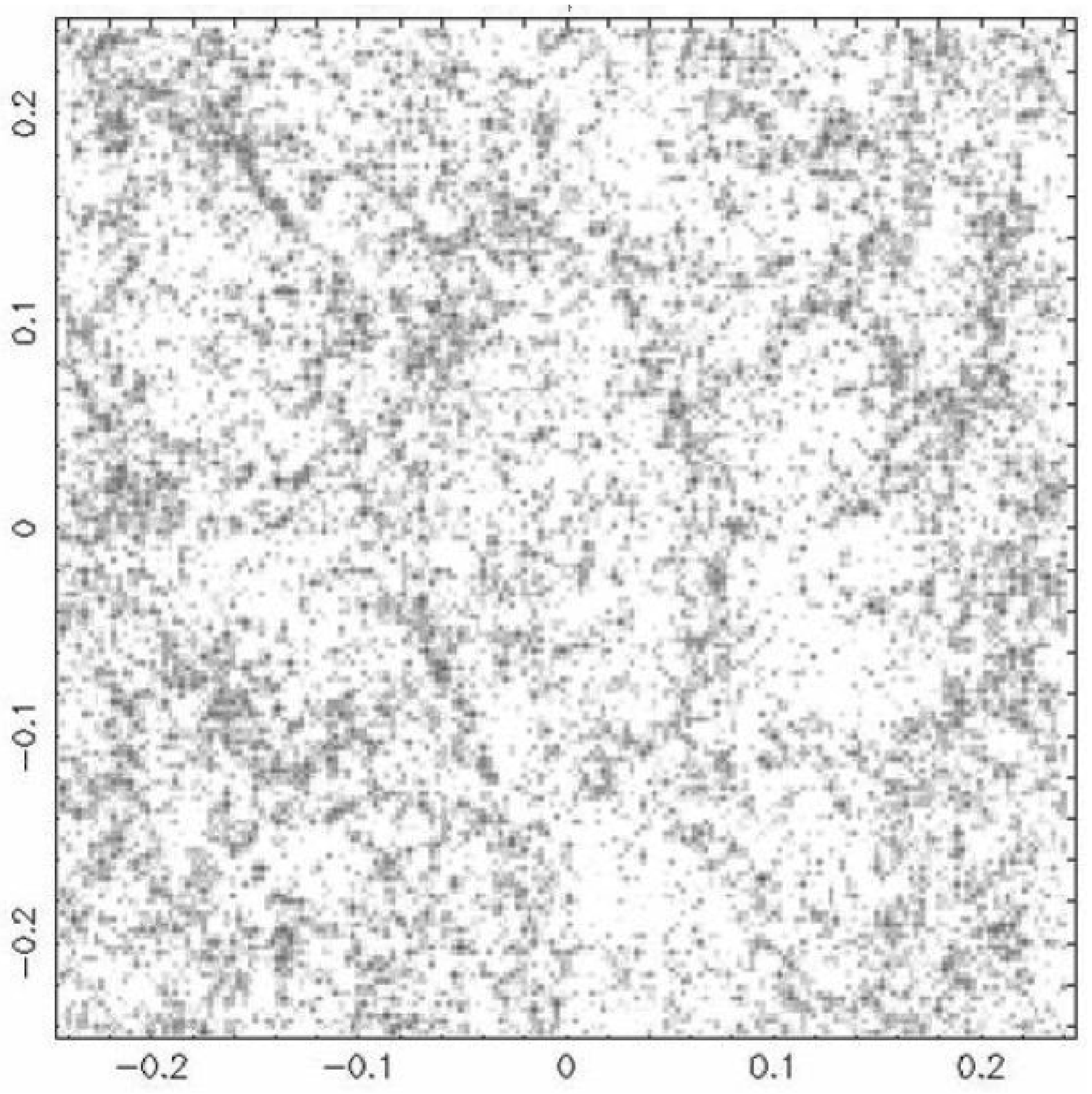}
\caption{ {\it (left)} A 1~square degree continuum simulation for the
SKA. The flux-density limit is 1~$\mu$Jy at 1.4~GHz (Jarvis et al. in
prep.). {\it (right)} 0.25~square degree HI simulation at a redshift of
$2.55 < z < 2.65$ (Obreschkow et al. in prep.) based on
the Millennium Simulation \citep{Springel05} and the semi-analytic
model of \cite{Croton06}.}\label{fig:jarvis_sim}
\end{figure}

As the name suggests the SKA will have a total collecting area of
$10^{6}$~m$^{2}$, making it $\sim 50$~times more sensitive than the EVLA. This increase in sensitivity enables the SKA
to push studies of galaxy formation and evolution to an unprecedented
level as the SKA will be able to detect the 21~cm transition of
neutral hydrogen in emission up to $z \sim 2$.  Thus, the SKA will
have the ability to trace the evolution of the HI mass function over
75 per cent of cosmic time (see e.g. Fig.~\ref{fig:jarvis_sim}). Even if this was conducted over relatively
small areas it would still provide a great leap in our knowledge of
galaxies and their dark matter haloes.  However, one of the big
challenges for the SKA is to couple this sensitivity to a large
field-of-view of order $\sim 50$~square degrees. This coupling of sensitivity
and field-of-view would make the SKA the most powerful survey
telescope in the world, with a survey speed around $10^{4}$ times
greater than the EVLA.

This means that within approximately one year the SKA could survey the
whole hemisphere in HI up to $z \sim 2$ with the expectation of
obtaining spectroscopic redshifts via 21~cm emission for $\sim 10^{9}$
galaxies. With such a survey the SKA would become the premier
instrument for measuring the evolution of Dark Energy via BAOs \citep{AR05}.

The SKA will also be able to conduct the deepest continuum
observations ever made in the radio regime. Fig.~\ref{fig:jarvis_sim} shows a simulated
square degree of sky at 1.4~GHz for a typical deep
(1$\mu$Jy) continuum survey with the SKA.  This shows the extent to which
the SKA will be able to probe galaxy formation and evolution up to the
highest redshifts, again like LOFAR the dominant sources are
starbursts and radio-quiet quasars. However, the sensitivity is such
that typical starburst galaxies can be detected up to and into the
epoch of reionisation. If the SKA reaches it's goal with respect to
the field-of-view then we expect to detect around $10^{7}$ galaxies
per 50~square degree pointing, in its deep continuum observations.

However, the SKA will not only be a redshift survey machine, the
science goals are broad and span everything from the search for life
on other planets through to the epoch of reionisation.

\section{The SKA Design Study}

The concept for the SKA has been around for almost two decades, with
the first mention of it in the literature in 1991 \citep{Swarup91,Wilkinson91,Noordam91,Braun91}. However, in the past few years this concept has gradually
turned into a real project. The SKA design study is now underway
across the world. This design study
covers the development of the technology needed to build
a SKA, however a critical component of the design study is the science
simulations which will dictate the final design of the SKA.

The science simulations are predominantly being carried out in five
Key Science areas. These are Cradle of Life, Probing the Dark Ages,
The origin and evolution of Cosmic Magnetism, Strong field tests of
gravity using pulsars and black holes and Galaxy evolution, cosmology and dark energy . In
these proceedings I have only described the science concerned with
galaxy formation and evolution.  In this section I briefly describe
the process by which the science simulations are being carried out.

\begin{figure}
\plotone{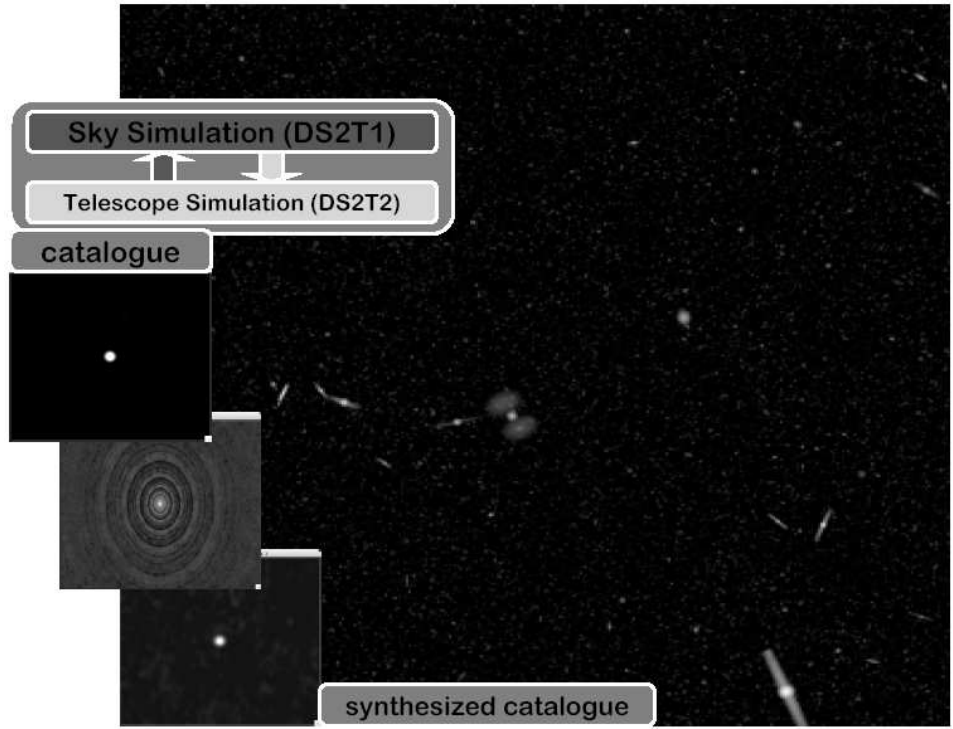}
\caption{Schematic representation of the SKA Design Study for the science simulations, describing how a `real sky' catalogue is put through a telescope simulator and the output is then passed back to the scientists to analyse. This is carried out for a range of SKA designs to optimise the telescope design.}\label{fig:jarvis_ds2}
\end{figure}

The science simulations can be split into two threads, the first is
concerned with both generating {\it real} skies (i.e. how the sky
looks if we had a perfect view of it) and also the analysis of data
after it has been processed through a simulated SKA telescope. This
middle step of taking real skies and putting them through a simulated
telescope is then carried out in the second thread. Thus, the tasks
enable scientists not familiar with radio astronomy techniques to have
an important input into the science simulations effort and also to realise the potential of such a powerful telescope.

A schematic representation of the interplay between the two threads
within the European Design Study,
called DS2-T1 (science simulations) and DS2-T2 (telescope
simulations), is shown in Fig.~\ref{fig:jarvis_ds2}. 
This process is now underway and will begin to produce important
information which will be fed into the overall design of the SKA (see
e.g. Fig.~\ref{fig:jarvis_sim}). These simulations will also be tested
to a certain extent with other SKA pathfinder telescopes operating at
higher frequencies than LOFAR, such telescopes currently under development are the
South African Karoo Array Telescope (KAT; http://www.ska.ac.za/kat) and the
Australian extended New Technology Demonstrator (xNTD;
http://www.atnf.csiro.au/SKA/xntd.html). These will both lead up to
the construction of the 10 per cent SKA planned for 2014, with
completion of the 100 per cent SKA in 2020.

\section{Summary}

I have discussed some of the extra-galactic science that will begin to
emanate from the new generation of radio telescopes, in particular
LOFAR and the SKA. However, both of these instruments will make great
strides in a wealth of other fields spanning astrophysics, cosmology
and particle physics and the reader is encouraged to read the
respective websites at (http://www.lofar.org and
http://www.skatelescope.org).

\acknowledgements I acknowledge funding from the SKA Design Study. I
would also like to thank a number of people who have contributed to
this article in a number of ways, these are Hans-Rainer Kl\"ockner,
 Lance Miller, Danail Obreschkow, Steve Rawlings, Huub
R\"ottgering and Richard Schilizzi.

\end{document}